# Hybrid plasmonic terahertz fibers for sensing applications

**Andrey Markov and Maksim Skorobogatiy**[*]

*Department of Engineering Physics, École Polytechnique de Montréal, C.P. 6079, succursale Centre-Ville Montreal, Québec H3C3A7, Canada*
*\* Corresponding author: maksim.skorobogatiy@polymtl.ca*

A novel plasmonic THz fiber featuring two metallic wires in a porous dielectric cladding is studied for resonant sensing applications. In our design, introduction of even lossless analytes into the fiber core leads to significant changes in the modal losses, which is used as a transduction mechanism.

Optical chemical and bio-sensors have attracted considerable attention in biochemistry and medicine where systems capable of highly sensitive, label-free and minimally invasive detection [1-2] are in high demand.

Most of the current THz sensors are realized in the non-resonant configurations where sample is interrogated directly by the THz light. In resonant sensors, changes in the sample properties are measured indirectly by studying variations in the optical properties of a resonant structure coupled to a sample. In resonant sensing method one typically employs resonant spectral transmission characteristics or photonic band-gap effect of sensing devices. The sensors detect shifts of the resonant wavelength in response to resonant structure changes caused by the presence of the analyte. These sensors include thin-film micro-strip lines resonators [3], photonic crystal waveguides [4], metallized periodic groove structures integrated into parallel-plate waveguides [5], thin metallic meshes [6], planar double split-ring resonator arrays [7], resonant cavities integrated into parallel-plate waveguides [8], and dielectric pipe waveguides [9]. In the amplitude-based detection methodology one operates at a fixed wavelength and records changes in the amplitude of a signal, which are then reinterpreted in terms of changes in the analyte refractive index. Among the advantages of this type of sensors are low cost and ease of fabrication, since no precisely engineered resonant or photonic band-gap structure is required.

In this Letter, we describe a novel refractometer based on practical plasmonic THz fibers that feature two metallic wires inserted into porous dielectric cladding [10, 11] [see Fig. 1(a)]. For the purpose of this paper, we refer to such fibers as composite fibers. In general, two-wire plasmonic THz waveguides combine both low loss performance and low group velocity dispersion. Moreover, the field distribution in the fundamental mode of a two-wire waveguide is similar to that emitted by the dipole-type photoconductive THz antennas. This allows efficient excitation of the two-wire waveguides using standard THz sources [10-12]. Finally, introduction of a porous cladding into the structure of a two-wire waveguide enables direct manipulations of such waveguides without the risk of perturbing the core-guided modes that are mostly confined in the gap between the wires.

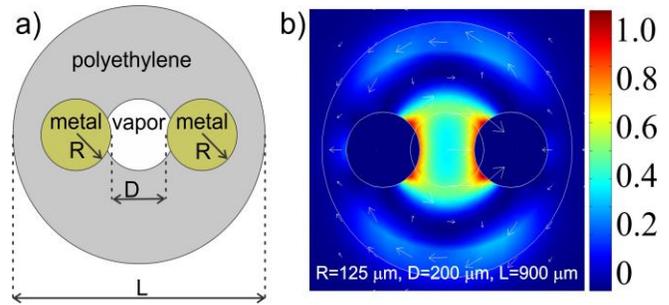

Fig. 1. a) Schematic of the composite three-hole fiber. b) Longitudinal flux distribution for the plasmonic mode in a composite fiber at 0.71 THz. Arrows show vectorial distribution of the corresponding transverse electric field.

The wave guiding in such composite fibers is most efficient for the light polarized parallel to the center line joining the two wires. A typical modal pattern represents the mixture of a plasmonic mode guided by the two wires, and a total internal reflection mode guided by the fiber plastic cladding. The presence of porous polyethylene cladding significantly complicates the modal structure of the waveguide. The modes of such fibers can be approximately described as "cladding" modes versus core-guided "plasmonic" modes. In order to distinguish predominantly plasmonic modes from the modes of a porous cladding, we first compare in Fig. 2 the dispersion relations, the absorption losses and the excitation efficiencies of the modes of a porous cladding alone (no wires) in the sub-1 THz frequency range. Solid colors in Fig. 2 define frequency ranges where modal excitation efficiency is higher than 5%, while dashed curves define spectral regions with less than 5% coupling efficiency from the Gaussian beam. In these simulations, we assume that the fiber center coincides with the focal point of a Gaussian beam.

Among all the modes of a composite fiber, there is one that clearly has no corresponding analogue among the modes of a stand-alone porous cladding, which we call the fundamental plasmonic mode of a composite fiber (blue color in Fig. 2). It extends into the very low frequencies (<0.1THz), while being well confined within the air-filled central hole of a fiber cladding at lower frequencies (up to 0.4 THz), as shown in Fig. 3 (a). In practical terms, it means that even at low frequencies the fundamental plasmonic mode is suitable for guiding THz light due to its strong confinement in the fiber hollow core and, as a consequence, relatively low loss (compared to the bulk absorption losses of polyethylene) and low group velocity dispersion, low sensitivity to bending, high tolerance to imperfections on the fiber surface and to perturbations in the environment. At higher frequencies (>0.4 THz) the modal power is displaced away from the central air hole [see Fig. 3 (a)] and into the polyethylene cladding surrounding the wires. Consequently, at higher frequencies absorption loss of the fundamental plasmonic mode approaches the bulk absorption loss of polyethylene.

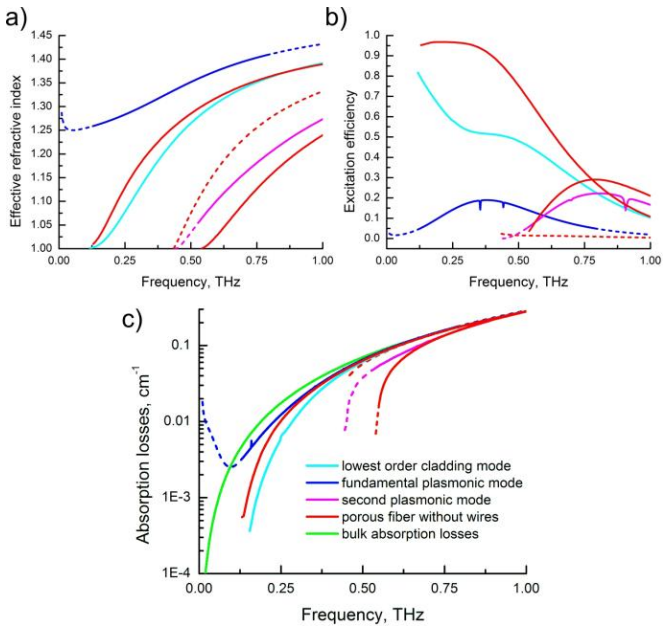

Fig. 2. a) Effective refractive indices, b) excitation efficiencies, and c) absorption losses for the various modes of a composite two-wire fiber shown in Fig. 1.

We now consider the lowest order cladding mode of a composite fiber in the 0.12 –1.00 THz frequency range (cyan color in Fig. 2). At lower frequencies (<0.3 THz) when operation wavelength is larger or comparable to the fiber size, the lowest order cladding mode has a strong presence outside of the fiber [see Fig. 3 (b)]. At higher frequencies, the modal fields tend to confine inside of the fiber cladding, thus resulting in the mode absorption losses similar to the bulk absorption losses of the cladding material. Note that the lowest order cladding mode of a composite fiber is in fact a hybrid mode that has a significant plasmon contribution. For this particular mode, the plasmon is propagating at the plastic/metal interface with almost no energy found in the central air hole. Optical properties of the modes of a porous fiber are presented in Fig. 2 in red color. We note that in the broad frequency range 0.13 – 1.0 THz, optical properties of the fundamental mode of a porous fiber are quite similar to those of the lowest order cladding mode of a composite fiber.

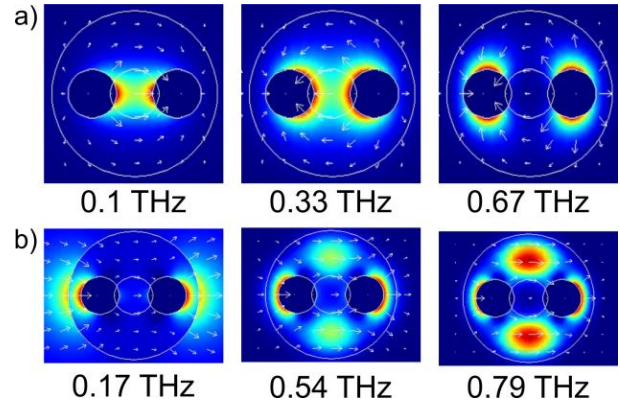

Fig. 3. Longitudinal flux distribution for the modes of a composite fiber a) fundamental plasmonic mode, b) lowest order cladding mode.

Finally, we consider the second order plasmonic mode of a composite fiber, which is presented in Fig. 2 in magenta color in the frequency range of 0.54 – 1.0 THz. From the corresponding field distributions showed in Fig. 4 it follows that the second order plasmonic mode is, in fact, a hybrid mode that has a significant plasmon contribution. Moreover, at lower frequencies (0.5 – 0.7 THz) the plasmon is propagating at the air/metal interface with a significant amount of energy concentrated in the central air hole between the two wires. In what follows we call this mode the second plasmonic mode as it can be used for low loss guidance of THz light within the central air hole of a composite fiber. At higher frequencies (>0.7 THz), the second plasmonic mode leaves the central air hole, while localizing in the vicinity of the metal/plastic/air junctions (see, for example, Fig. 4, 0.79 THz). This results in significant energy transfer into the cladding, and, consequently, modal absorption losses of this mode become comparable to the bulk absorption losses of a polyethylene cladding.

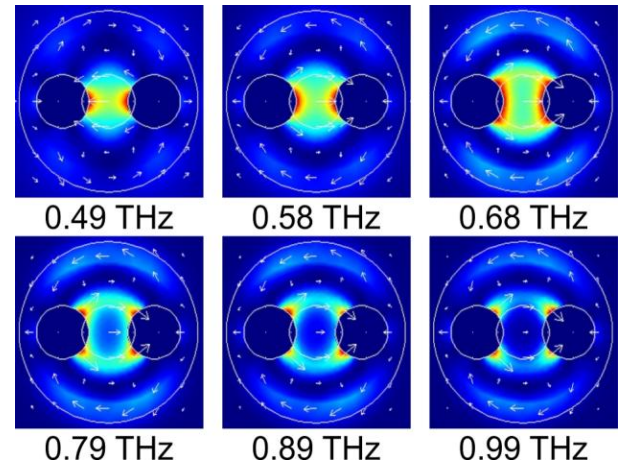

Fig. 4. Longitudinal flux distribution for the second plasmonic mode of a composite fiber at various frequencies.

For the sensing applications we suggest using the second plasmonic mode of the composite fiber due to the most rapid changes of the fiber optical properties with frequency. Also, it is natural to expect high sensitivity of such sensors due to an almost perfect overlap between analyte filling the hollow core and the optical mode of a sensor. Transverse distribution of the longitudinal power flux of such mode at 0.71THz is shown in Fig. 1 (b). The choice of this frequency is justified by the fact that in its vicinity, localization of the mode changes from the core bound to cladding bound. As we will see in the following, such behavior is advantageous for sensing applications.

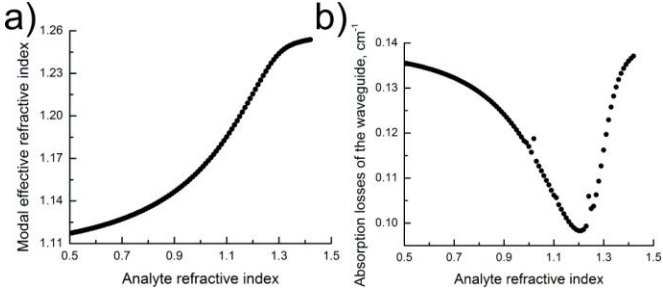

Fig. 5. Changes in the optical properties of a plasmonic mode as a function of the analyte refractive index. a) Modal effective refractive index, b) absorption losses.

To demonstrate sensing applications we first assume that the central hole of the fiber is filled with analyte in gaseous or aerosol forms. In this case, the core refractive index is changed uniformly and throughout the fiber core. We suppose that the loss of a fiber cladding material is 0.14 cm$^{-1}$, which corresponds to the bulk material loss of polyethylene at 0.71 THz [13]. In our simulations we only vary the real part of the analyte refractive index, while assuming that the gaseous analyte has negligible absorption. This is only to show the resonant nature of the transduction mechanism. Unequivocally, the same sensor will also detect changes in the analyte absorption.

The outer diameter of the composite fiber in Fig. 1 is 900 µm, the wire diameter is 250 µm, the air hole diameter is equal to the wire diameter and the gap between the wires is 200 µm. In Fig. 5 we show how optical properties of the plasmonic mode change when varying the analyte refractive index (core refractive index). In these simulations we kept the operation frequency fixed at 0.71 THz. Clearly, strong changes in the fiber optical properties with respect to changes in the analyte refractive index are desirable for the optimal design of a fiber refractometer. From Fig. 5(b) we note, for example, that the rate of change in the modal absorption loss is the highest when the core refractive index is close to 1, which is most suitable for the design of refractiometers operating with gaseous analytes. This high sensitivity of the modal loss on the value of the core refractive index is directly related to the choice of the operational frequency of 0.71 THz at which core guided mode shows significant changes in its localization preference from the hollow core at low frequencies into the plastic cladding at higher frequencies.

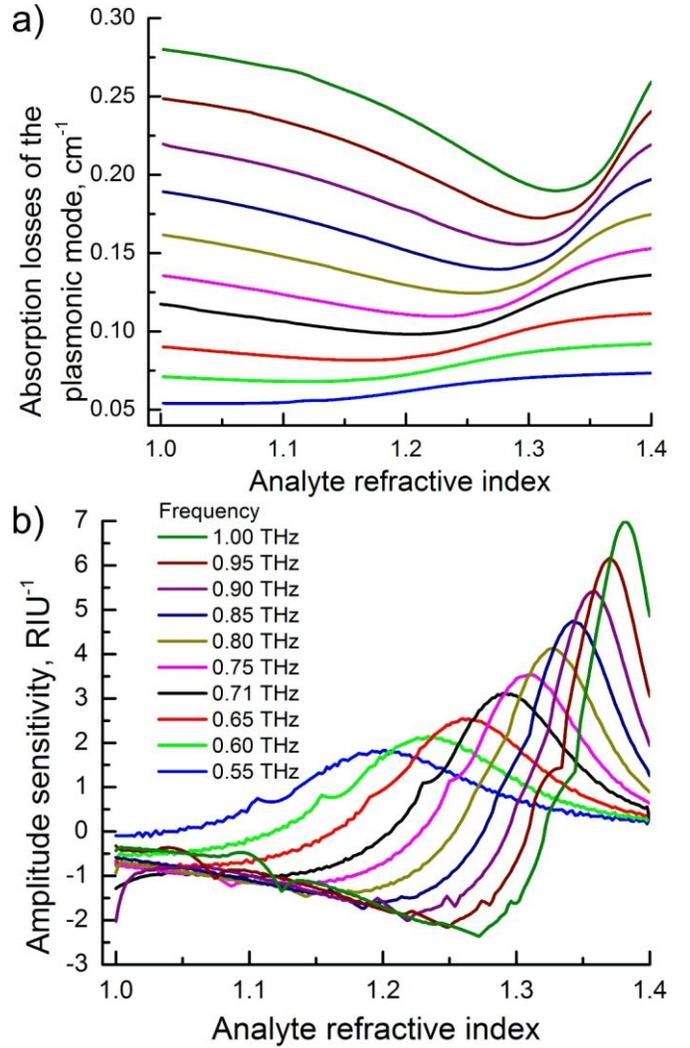

Fig. 6. a) Absorption losses of the plasmonic mode, and corresponding b) Sensitivity of the refractometer as a function the refractive index of the analyte at various values of the operation frequency. The wiggles correspond to the anticrossings of the plasmonic modes with fiber cladding modes

We now consider various factors that influence resolution of the refractometer. In the following simulations we use 1.514 as a frequency independent refractive index of the polyethylene cladding with frequency dependent material loss $\alpha\left[cm^{-1}\right]=0.28\cdot\nu^{2}$ ($\nu$ is in THz) [13]. As before, we suppose that analyte is lossless.

First of all, form Fig. 6 we notice that for a given value of the analyte refractive index $n_a$, sensor's sensitivity can be optimized by the judicial choice of the operation frequency $\nu$. Particularly, in Fig. 6(a) we present absorption losses of the plasmonic mode of a three-hole composite fiber as a function of the analyte refractive index for various choices of the operation frequency. As demonstrated in [14], for a given value of the analyte refractive index, the optimal choice of the operation frequency is the one that maximizes an amplitude sensitivity defined as

$$S_a(n_a,\nu)=\left(\partial\alpha_m(n_a,\nu)/\partial n_a\right)/\alpha_m(n_a,\nu) \qquad (1),$$

where $\alpha_m(n_a,\nu)$ is the absorption loss of a plasmonic mode. Resolution of a sensor can then be calculated by assuming that 1% change in the transmitted amplitude can be reliably detected, from which sensor resolution is calculated as $0.01/S_a(n_a,\nu)$. Consider, for example, gaseous analytes with refractive index close to 1. From Fig. 4 we observe that the nature of modal localization changes rapidly in the 0.7-0.9 THz frequency range from the core-guided to the cladding-guided. Consequently, this results in significant changes in the fiber absorption losses in this frequency range. Therefore, it is not surprising to find that the refractometer's sensitivity achieves its maximal value of $2.02\,RIU^{-1}$ at 0.9 THz, with the corresponds resolution of $5\cdot 10^{-3}\,RIU$. The maximal sensitivity is achieved at 1.0 THz and is equal to $6.98\,RIU^{-1}$, resulting in the sensor resolution of $1.4\cdot 10^{-3}\,RIU$.

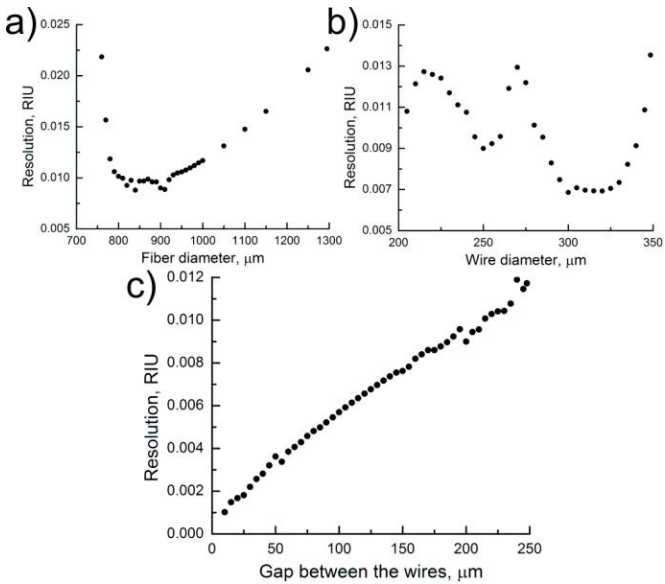

Fig. 7. Resolution of the refractometer as a function of a) fiber diameter, b) wire diameter, c) gap between the wires.

The geometric parameters of the fiber can be also optimized to increase sensor resolution. Thus, Fig. 7 presents resolution of the refractometer for different values of the fiber diameter, the wire diameter and the gap size between the wires. In these simulations we kept the operation frequency fixed at 0.71 THz and the analyte refractive index close to 1. Since plasmonic mode of a composite fiber is generally well confined within the central hole region of the fiber, the refractometer resolution is not strongly sensitive on the fiber and wire diameters [see Figs. 7 (a), (b)]. In contrast, inter-wire gap size has the strongest influence on the refractometer resolution [see Fig. 7 (c)]. There is virtually a linear dependence of the resolution on the distance between the wires. Smaller gaps between the wires result in faster (with frequency or core refractive index) changes in the nature of modal localization and, thus, result in higher sensitivities. Thus, resolution as small as $3\cdot 10^{-3}\,RIU$ can be achieved for the inter-wire distance of 50 μm.

In conclusion, we have proposed a novel refractometer based on practical plasmonic THz fibers that feature two metallic wires inserted into porous dielectric cladding. Introduction of even lossless analytes into the fiber core leads to significant changes in the modal losses, which is used as a transduction mechanism. Resolution of the refractometer has been investigated numerically as a function of the operation frequency and the geometric parameters of the fiber. We have shown that amplitude-based detection method leads to sensitivities to the changes in the gaseous analyte refractive index on the order of ~$10^{-3}$ RIU.

### References

1. N N. Laman, S. Harsha, D. Grischkowsky, and J. Melinger, Biophys. J. **94**, 1010-1020 (2008).
2. C. Markos, W. Yuan, K. Vlachos, G. E. Town, and O. Bang, Opt. Express **19**, 7790-7798 (2011).
3. M. Nagel, P. Haring Bolivar, M. Brucherseifer, H. Kurz, A. Bosserhoff, and R. Buttner, Appl. Phys. Lett. **80**, 154-156 (2002).
4. Hamza Kurt and D. S. Citrin, Appl. Phys. Lett. **87**, 241119 (2005).
5. M Nagel, P Haring Bolivar and H Kurz, Semicond. Sci. Technol. **20**, S281-S285 (2005).
6. H. Yoshida, Y. Ogawa, Y. Kawai, S. Hayashi, A. Hayashi, C. Otani, E. Kato, F. Miyamaru, and K. Kawase, Appl. Phys. Lett. **91**, 253901 (2007).
7. J. O'Hara, R. Singh, I. Brener, E. Smirnova, J. Han, A. Taylor, and W. Zhang, Opt. Express **16**, 1786-1795 (2008).
8. Rajind Mendis, Victoria Astley, Jingbo Liu, and Daniel M. Mittleman, Appl. Phys. Lett. **95**, 171113 (2009).
9. B. You, J. Lu, C. Yu, T. Liu, and J. Peng, Opt. Express **20**, 5858-5866 (2012).
10. A. Markov, M. Skorobogatiy, "Two-wire Terahertz Fibers with Porous Dielectric Support," arXiv:1303.2529
11. A. Markov and M. Skorobogatiy, Opt. Express, **21**, 12728-12743 (2013).
12. M. Mbonye, R. Mendis, and D. Mittleman, Appl. Phys. Lett. **95**, 233506 (2009).
13. A. Markov, A. Mazhorova, M. Skorobogatiy, IEEE Trans. Terahertz Sci. Technol., **3**, 96-102 (2013).
14. M. Skorobogatiy, "Resonant bio-chemical sensors based on Photonic Bandgap waveguides and fibers," in *Optical guided-wave Chemical and Biosensors II*, M. Zourob, A. Lakhtakia,(Springer-Verlag, 2010), pp. 43-72.

**References**

[1] N. Laman, S. Harsha, D. Grischkowsky, and J. Melinger, "High-resolution waveguide THz spectroscopy of biological molecules," Biophysics Journal **94**(3), 1010–1020 (2008).

[2] C. Markos, W. Yuan, K. Vlachos, G. E. Town, and O. Bang, "Label-free biosensing with high sensitivity in dualcore microstructured polymer optical fibers," Optics Express **19**(8), 7790–7798 (2011).

[3] M. Nagel, P. Haring Bolivar, M. Brucherseifer, H. Kurz, A. Bosserhoff, and R. Buttner ,"Integrated THz technology for label-free genetic diagnostics," Applied Physics Letters **80**(1), 154-156 (2002).

[4] Hamza Kurt and D. S. Citrin, "Coupled-resonator optical waveguides for biochemical sensing of nanoliter volumes of analyte in the terahertz region," Applied Physics Letters **87**, 241119 (2005).

[5] M Nagel, P Haring Bolivar and H Kurz, "Modular parallel-plate THz components for cost-efficient biosensing systems," Semiconductor Science and Technology **20**, S281-S285 (2005).

[6] H. Yoshida, Y. Ogawa, Y. Kawai, S. Hayashi, A. Hayashi, C. Otani, E. Kato, F. Miyamaru, and K. Kawase , "Terahertz sensing method for protein detection using a thin metallic mesh," Applied Physics Letters **91**, 253901 (2007).

[7] J. O'Hara, R. Singh, I. Brener, E. Smirnova, J. Han, A. Taylor, and W. Zhang, "Thin-film sensing with planar terahertz metamaterials: sensitivity and limitations," Optics Express **16**(3), 1786-1795 (2008).

[8] Rajind Mendis, Victoria Astley, Jingbo Liu, and Daniel M. Mittleman, "Terahertz microfluidic sensor based on a parallel-plate waveguide resonant cavity," Appl. Phys. Lett. **95**, 171113 (2009).

[9] B. You, J. Lu, C. Yu, T. Liu, and J. Peng, "Terahertz refractive index sensors using dielectric pipe waveguides," Opt. Express **20**(6), 5858-5866 (2012).

[10] A. Markov, M. Skorobogatiy, "Two-wire Terahertz Fibers with Porous Dielectric Support," arXiv:1303.2529

[11] A. Markov and M. Skorobogatiy, "Two-wire terahertz fibers with porous dielectric support," Optics Express, **21**(10), 12728-12743 (2013).

[12] M. Mbonye, R. Mendis, and D. Mittleman, "A terahertz two-wire waveguide with low bending loss," Applied Physics Letters **95**, 233506 (2009).

[13] A. Markov, A. Mazhorova, M. Skorobogatiy, "Planar porous THz waveguides for low-loss guidance and sensing applications," IEEE Transactions on Terahertz Science and Technology, **3**(1), pp. 96-102 (2013).

[14] M. Skorobogatiy, Resonant bio-chemical sensors based on Photonic Bandgap waveguides and fibers. "Optical guided-wave Chemical and Biosensors II," (M. Zourob, A. Lakhtakia, editors, Springer-Verlag 2010), pp. 43-72.